# Quantum Computing Using Electrons Floating on Liquid Helium

M. I. Dykman[1] and P. M. Platzman[2]

[1] Michigan State University, Dept. of Physics & Astronomy, East Lansing, MI 48824

[2] Bell Laboratories, Lucent Technologies, Murray Hill, NJ 07974

**Abstract**

The system of electrons trapped in vacuum above the liquid helium surface displays the highest mobilities known in condensed matter physics. We provide a brief summary of the experimental and theoretical results obtained for this system. We then show that a quasi-2D set of $N > 10^8$ electrons in vacuum trapped in 1D hydrogenic levels above a micron-thick helium film can be used as an easily manipulated strongly interacting set of quantum bits. Individual electrons are laterally confined by micron sized metal pads below the helium. Information is stored in the lowest hydrogenic levels. Using electric fields at temperatures of $10^{-2}$ K, changes in the wave function can be made in nanoseconds. Wave function coherence times are .1 millisecond. The wave function is read out using an inverted dc voltage which releases excited electrons from the surface, or using SETs attached to the metal pads which control the electrons.

## 1. Introduction

Constructing an Analog Quantum Computer (AQC) is an interesting and challenging problem. It requires creating a very unusual system, which should be composed of many ($>10^2$) individual quantum objects. They should interact with each other, but their coupling to the outside world must be reduced to an extremely low level. In order to perform a computation, one must be able to manipulate the individual objects. This includes changing their states and their interactions with each other. Finally, it is necessary to be able to "read out" some properties of the quantum system, at some time which signals the end of a computation. To date, research has been largely focussed on the individual "building blocks" of the AQC, which are two-state systems and can be associated with "quantum" bits, qubits, although in principle systems with a larger number of states would be acceptable as well provided one could control them.

The problem of creating a quantum computer is extremely complicated, because it is necessary to meet all the conditions listed above and to be able to manipulate individual qubits on the time scale small compared to the characteristic time over which they loose quantum coherence. Many groups are currently working on this problem, and several types of systems have been suggested, including atoms in traps,[1,2,3,4] cavity quantum electrodynamics systems,[5] bulk NMR systems,[6,7,8] quantum dots,[9,10,11,12] nuclear spins of atoms in doped silicon devices,[13] localized electron spins in semiconductor heterostructures,[14] and Josephson-junction based systems,[15,16,17,18] and in a number of cases proof of the principle has been demonstrated. However, for all these suggestions there are high technological and scientific barriers which must be overcome to create a system which will then become a useful quantum computer.

We believe, based on realistic estimates,[19] that there already exists a system, which is a good candidate for a scalable quantum computer that has an easily manipulated set of qubits with extremely long coherence times. This is a system of electrons floating on the surface of superfluid helium at very low temperatures. Such system is incredibly clean, with the *highest* electron mobility ever achieved in a condensed matter system.[20] The electrons can function as qubits, provided they are confined by micron size electrodes below the helium surface. We showed[19] that, under realistically obtainable conditions, i.e.

correct geometry, temperature, magnetic field, etc, the system of quasi 2D electrons can behave as a many-qubit AQC. In addition to the suggested "system design", we calculated the very important decoherence effects to show that they were acceptably small.

The purpose of this paper is to discuss in more detail the idea of AQC based on electrons on helium, particularly since the system is a somewhat unfamiliar one. Compared to Reference 19, here we will give a fuller account albeit brief of what is already known about the relevant physics of many unconfined 2D electrons. We will then discuss the new physics connected with electrons confined laterally on micron thick films, and how they can be used as qubits.

First we will present results relevant to a single qubit. In particular, we will analyze decoherence times at finite temperature and show how we may manipulate them by suppressing unwanted lateral degrees of freedom. We will then consider in some detail the manipulation of two qubits, i.e. a two-qubit gate. Finally we will allude to many qubits. One possible readout process will also be analyzed. We conclude this review with a discussion of how this system may be used to study several fundamentally important physical problems, which cannot be solved in a conceivable time with ordinary computers.

## 2. Electrons on helium

### 2.1. The electron energy spectrum

The standard experimental geometry and the energy spectrum of electrons floating on the helium surface are shown schematically in Fig. 1. The helium partially fills a parallel plate capacitor. The surface is charged using a filament (a cathode). The capacitor has a voltage across it, which for a fixed geometry determines the charge density and provides a uniform neutralizing background for the electrons.

An electron on a bulk helium film of thickness $d \geq .5$ μm is trapped primarily by a comparatively weak image potential, which to a good approximation is of the form $V(z) = -\Lambda e^2/z$ (Fig. 1)[21] where $z$ is the direction normal to the surface. In this case $\Lambda = (\varepsilon - 1)/4(\varepsilon + 1) \cong .01$ [the dielectric constant of liquid helium $\varepsilon \cong 1.057$]. Because there is a barrier of 1eV for penetration into the helium, the electrons $z$-motion is quantized. The corresponding energy spectrum is 1D hydrogenic, that is the $m^{th}$ state has an energy $E_m = -R / m^2$, with an effective Rydberg energy $R = \Lambda^2 e^4 m_e / 2\hbar^2 \approx 8K \approx$ 160 GHz, and effective Bohr radius $r_B = \hbar^2 / m_e e^2 \Lambda \approx 76 \overset{\circ}{A}$ [22] [here and below we give frequencies and energies in the units of temperature, which corresponds to $k_B = \hbar = 1$].

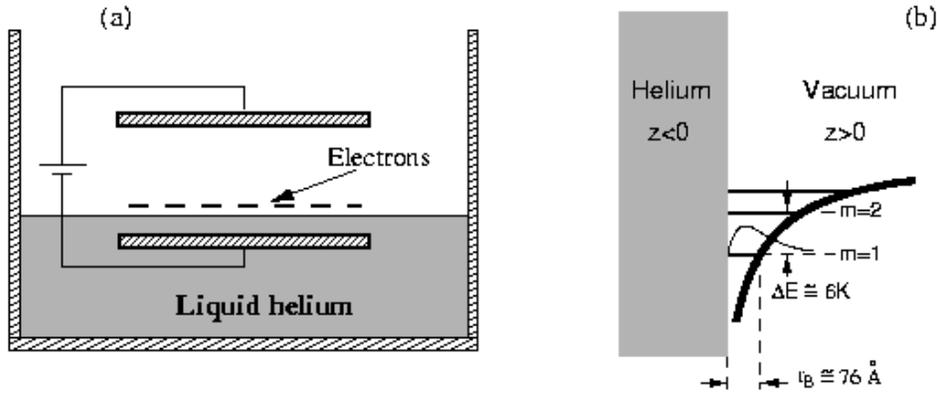

Fig.1.(a) The geometry of experiments on a spatially uniform electron system on helium surface (schematically). Electrons are trapped at the helium vacuum interface. A capacitor determines the maximal electron density. It also allows to apply an electric field $\mathcal{E}_\perp$ which presses electrons to the surface. (b) The hydrogenic energy levels of motion transverse to the helium surface, along with a typical $m = 1$ wavefunction, displayed in the image potential $V(z) = -\Lambda e^2 / z$ for $z > 0$.

In 1976 Grimes et al.[22] reported a spectroscopic study of electronic transitions between the energy levels of motion transverse to helium surface. The level spacing was controlled by a static electric field $\mathcal{E}_\perp$ normal to the helium surface, with a potential $V_s = e\mathcal{E}_\perp z$, which Stark-shifts the energy levels. The transitions were observed at frequencies from 130 to 220 GHz by measuring the microwave absorption derivative at a fixed frequency as the splitting between the states was tuned past resonance by modulating the field $\mathcal{E}_\perp$, see Fig. 2. The calculated linear Stark tuning rates are appreciable, i.e 0.3 GHz and 1.1 GHz per 1V/cm for the $m = 1$ and $m = 2$ states, respectively. This means that the separation between the two lowest levels in Fig. 1 can be tuned by modest dc field at the rate of ~1GHz/(V/cm). This will be important for our AQC. The good agreement between the calculated shifts and experiments is seen from the insert of Fig. 2.

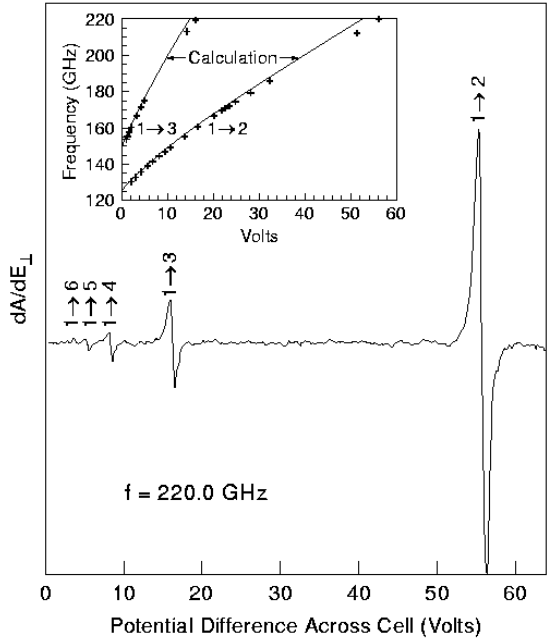

Fig. 2 . Experimental recording of mm-wave absorption derivative vs. the potential difference across the experimental cell taken at frequency 220 GHz and temperature 1.2 K [22]. The Stark effect brings the frequencies of the transitions from the ground ($m = 1$ in Fig. 1) to the excited states ($m = 2,3,…$) in resonance with the applied radiation. Inset: the dependence of the transition frequencies on the applied field. The crosses are measured data points while the solid curves are the result of the calculation.

The linewidth of the resonant absorption lines observed in the experiment[22] was ~1 GHz, i.e. the decoherence rate was quite high, about four-five orders too large to allow us to use such transitions as qubits. However, at the temperatures of the experiment, $T > 1.2$K, this rate was determined by electron scattering by He atoms in the vapor. The vapor density is exponential in temperature. In the temperature range from 1.2 to 0.7 K it decreases by 3 orders of magnitude, and below 0.1 K becomes essentially equal to zero.

## 2.2. Electron-ripplon scattering

For $T < 0.7$K, the only significant electron coupling to the outside world is to thermally excited height variations $\delta(\mathbf{r},t)$ of the helium surface, where $\mathbf{r}$ is the electron in-plane coordinate. The height variations of the surface are described as propagating capillary waves having a dispersion relation $\omega^2(k) = gk + (\sigma/\rho)k^3$, [21] where $\rho = 0.145$g/cm$^3$ and $\sigma = 0.37$erg/cm$^2$ are the mass density and surface tension of the superfluid He$^4$, respectively, and $g$ is the gravitational constant. The electron-ripplon coupling Hamiltonian is

$$H_{er} = e\hat{\mathcal{E}}_T \delta ,$$   (1)

where $\hat{\mathcal{E}}_T$ is an operator characterizing the effective $z$-directed electric field on the electron. It is given by the sum of the external field $\mathcal{E}_\perp$ and the field which comes from the image potential due to the surface distortion. This image field is nonlocal in $\mathbf{r}$. This means that the image part of (1) is an integral over $\mathbf{r}'$ of $f(\mathbf{r}-\mathbf{r}')\,\delta(\mathbf{r}')$, with the function $f$ depending also on the out-of-plane electron coordinate $z$. To the order of magnitude, this field is $\sim 10^2 - 10^3$ V/cm.

The range of the ripplon wave vectors $\mathbf{k}$ which are important, depends on the lateral (in-plane) length scales of interest. For a free electron typical $k$'s are twice the thermal de Broglie wave vector $k_T = \sqrt{2m_e\,k_B T/\hbar}$. For $T = 10^{-2}$ K  $k_T \cong 10^5\,\mathrm{cm}^{-1}$, which means that the gravity term is not important in the ripplon dispersion law. As we will see below, in the suggested AQC geometry, for electrons confined by micro dots again the typical $k$ are $<\sim 5 \times 10^5\,\mathrm{cm}^{-1}$. The typical ripplon frequencies are $\omega_r = \omega(k_{\max}) < 4 \times 10^{-3}$K. Therefore even at $10^{-2}$K many ripplons are present, and the mean square displacement of the surface is determined by thermal fluctuations, having a root mean square value of

$$\delta_T \equiv \left(\frac{k_B T}{\sigma}\right)^{1/2} \cong 2 \times 10^{-9}\,\mathrm{cm}.$$  In our temperature range it is the coupling to ripplons which leads to most of the important decoherence effects.

From the point of view of transitions between the hydrogenic states, which are of interest for AQC, two types of decoherence processes are important: ripplon-induced scattering between different hydrogenic states (more precisely, between the bands of 2D in-plane motion corresponding to the different states $m = 1,2,\ldots$ in Fig. 1), which gives rise to a finite lifetime of an electron in the excited band $T_1$, and intraband processes which give rise to breaking of the phase difference between the states in the different bands. Because ripplon energies are so small, the scattering by ripplons is essentially elastic. Therefore interband $m = 2 \rightarrow m = 1$ scattering requires a ripplon with a wave vectors $k \sim r_B^{-1}$  i.e., $\hbar^2 k^2 / 2m_e \cong E_2 - E_1$. A simple calculation of this $T_1$ process shows that,

$$\frac{\hbar}{T_1} \cong R \times \left(\frac{\delta_T}{r_B}\right)^2 \qquad . \tag{2}$$

Since $\left(\delta_T / r_B\right) \cong 10^{-3}$ for temperature 10 mK, the reciprocal lifetime is $10^{-6}$ of the transition frequency $\approx 120$ GHz.

The dephasing rate is essentially determined by in-plane scattering, in which the electron does not make an interband transition. For transitions $1 \rightarrow 2$, the corresponding coupling $H_d = <1|H_{er}|1> - <2|H_{er}|2>$ is determined by the difference of the *diagonal* matrix elements of the operator $H_{er}$ on the wave functions for the in-plane motion in the states $m = 1,2$.[23] This rate, which is also temperature dependent, is smaller than the transport scattering rate (reciprocal momentum relaxation time $\tau^{-1}$) for in-plane motion in the ground state $m$=1. However, for weak pressing fields $\mathcal{E}_\perp$ they are of the same order of magnitude.

In the single-electron approximation the relaxation rate of the in-plane momentum due to quasielastic scattering by ripplons can be analyzed in a standard way giving $\tau^{-1} \cong e^2 \mathcal{E}_T^2 / 4\sigma$. Here, $\mathcal{E}_T$ is a temperature dependent electric field determined by the square root of the appropriately weighted and averaged squared matrix element $\left|<1|H_{er}|1>\right|$. The corresponding scattering rates have been carefully tested experimentally by measuring the in-plane electron mobility at low frequencies,[24] and good agreement has been found between the theory and experiment[21] in a broad range of electron densities and temperatures.

From the theoretical results one expects to have extremely small relaxation rate for low temperatures. Indeed, if one uses a standard expression which relates the mobility and $\tau^{-1}$, the results of the experiment[20] give the record-low value $\tau^{-1} \approx 10^7$ s$^{-1}$ for 0.1K. In fact, this value is not very far from the upper bound on the homogeneous linewidth of 15 MHz for the $1 \rightarrow 2$ transitions, obtained by Volodin and Edel'man for 0.4K.[25]

In-plane electron-ripplon scattering becomes more complicated in the situation which, as we shall see, is of interest to the AQC when a magnetic field $B_\perp$ is applied normal to the helium surface.[26,27] This geometry is, of course, the one used to observe the quantum Hall effect. However since the electron system on bulk helium is non-degenerate, there is no quantum Hall effect. Nevertheless the field $B_\perp$ changes the electron motion dramatically. Neglecting electron-electron interaction, the energy spectrum of in-plane electron motion becomes a set of discrete Landau levels, with spacing $\cong \hbar \omega_c$, where $\omega_c = eB_\perp / m_e c$ is the cyclotron frequency. This means that quasi-elastic scattering by a single ripplon is forbidden. For weak coupling, the dephasing rate, it turns out, is determined by two-ripplon processes in which a ripplon is scattered by an electron into another ripplon with nearly the same energy, and as a result the phase of the electron wave function is changed.[28] In contrast to the case where there is no magnetic field, this dephasing rate is of the fourth rather than second order in the electron-ripplon coupling constant. We will analyze this scattering in more detail for confined qubits later.

### 2.3. Many-electron effects

In all experiments performed on electrons on bulk helium,[21] the electron density $n$ was small in the sense that the electron system was nondegenerate, with the Fermi energy $\varepsilon_F = \pi \hbar^2 n / m_e \leq 40\,\mathrm{mK}$. Yet $n$ was large in the sense that the energy of the electron-electron interaction exceeded the electron thermal kinetic energy, i.e.

$$\Gamma = e^2 (\pi n)^{1/2} / k_B T \gg 1 \,. \tag{3}$$

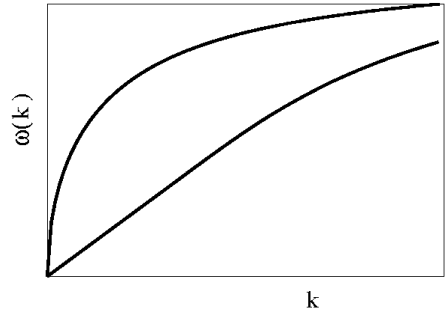

Fig. 3. Long-wavelength ($k \ll \sqrt{n}$) frequency spectrum of a 2D Wigner solid (schematically). The acoustic branch is a shear mode, whereas longitudinal phonons have the dispersion law $\omega(k) = \omega_p \left( k / \sqrt{n} \right)^{1/2}$, with $\omega_p = (2\pi e^2 n^{3/2}/m_e)^{1/2}$. If a magnetic field is applied normal to the helium surface, the spectrum changes into the magneto-plasma mode which starts from the cyclotron frequency $\omega_c$ and a low-frequency mode with $\omega(k) = (\omega_p^2 / \omega_c) \left( k / \sqrt{n} \right)^{3/2}$. If the electron system is confined, as in the proposed AQC, there emerges a gap at zero frequency.

One of the most striking effects of the electron-electron interaction is Wigner crystallization, in which electrons minimize the energy of mutual repulsion by forming a crystal. The crystallization occurs for $\Gamma \cong 130$, and its features and consequences continue to attract much attention.[21] The classical 2D Wigner solid (i.e., for $k_B T >> \varepsilon_F$) forms a triangular lattice. In the absence of coupling to ripplon excitations it has a rather simple harmonic phonon spectrum,[29] the long-wavelength part of which is shown in Figure 3. The lowest branch, linear in wave vector is the shear mode, while the upper branch $\omega \sim k^{1/2}$ is the longitudal plasmon mode.

In a beautiful experiment, Grimes and Adams[30] found that, below a certain temperature (0.457 K for an electron density $n = 4.5 \times 10^8 \text{cm}^{-1}$), the in-plane microwave conductivity as a function of frequency showed a complicated pattern of resonances (see insert in Fig. 4). These resonances were connected quantitatively by Fisher, Halperin and Plaztman[31] with the onset of long-range order in the electron system, i.e. their appearance allowed one to trace experimentally the phase transition line shown in Fig. 4.[30]

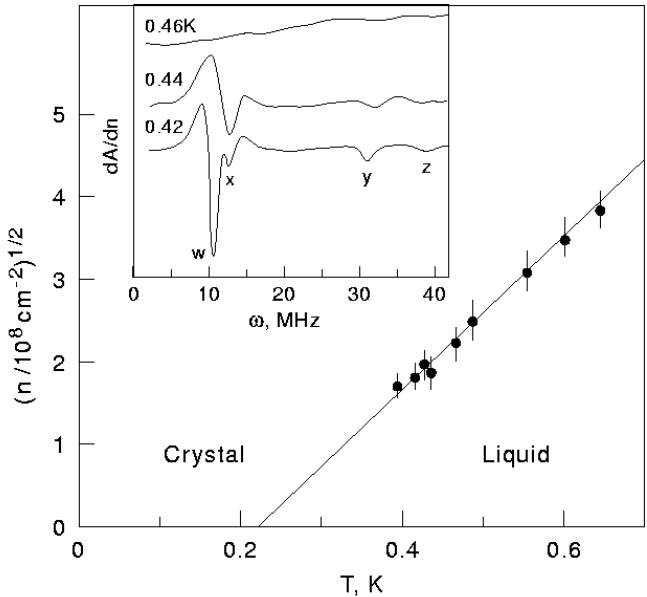

Fig.4. Portion of the solid-liquid phase boundary for a classical 2D sheet of electrons on liquid helium[30]. The data points denote the melting temperatures measured at various electron densities from the sudden appearance of resonant electron absorption at radio-frequencies, as shown in the inset, which is due to coupled plasmon-ripplon resonances (see Fig. 5). The solid line corresponds to $\Gamma = 137$.

The idea of the theory[31] is that, because of the known $\left(e \hat{\mathcal{E}}_T \delta\right)$ electron-ripplon coupling Eq. (1), longitudinal phonons of the Wigner crystal are coupled to ripplons. When there is crystalline order, a phonon in the Wigner crystal with the wave vector **k** is coupled to ripplons with the wave vectors **k** + **G** with different reciprocal lattice vectors **G**. In other words, phonon-ripplon scattering is accompanied by Umklapp processes in the crystal. The calculated dispersion law of the resulting coupled modes is shown in Figure 5. The positions of the resonances observed[30] and shown in Figure 5 are determined by this dispersion law and by the values of the wave vectors in the cavity used in the experiment. There is excellent agreement between theory and experiment.

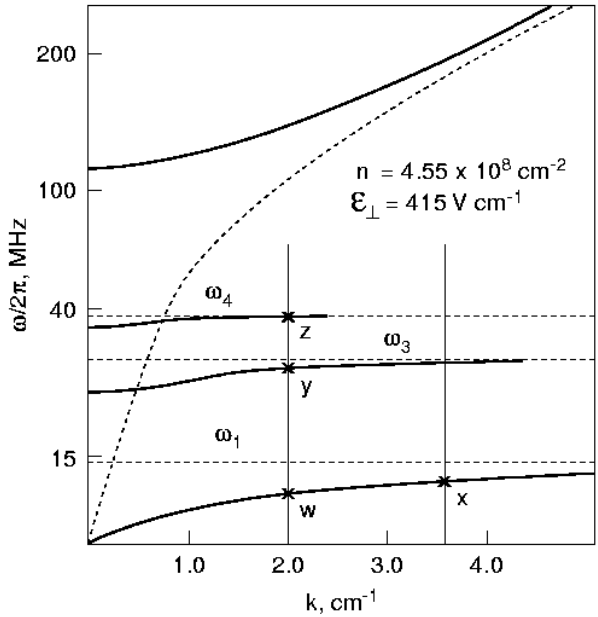

Fig. 5. Schematic of the dispersion relation of the longitudinal coupled modes of the Wigner solid on helium surface[31]. The dashed lines show the uncoupled longitudinal mode (cf. Fig. 3) and the ripplon modes with the wave numbers which correspond to the reciprocal lattice vectors of the Wigner solid. The vertical lines represent the wave vectors excited in the experiment[30], and the stars show the resonant frequencies observed in this experiment. The resonances are labeled as in the inset of Fig. 4.

Even where electrons do not form a Wigner crystal, the electron-electron interaction is still strong since $\Gamma \gg 1$. The electron system remains a strongly correlated non-degenerate electron fluid. In such a case one might expect that the electron-electron interaction will significantly change the in-plane mobility. However, as we showed[32] for a broad range of temperatures and electron densities, the single-electron theory is essentially correct in the absence of a magnetic field. This happens because the duration of electron-ripplon collisions is small, $\cong \hbar (k_B T)^{-1}$, and the interaction with other electrons only slightly changes the electron motion during that time.

The situation becomes completely different in the presence of the magnetic field $B_\perp$. Here the electron-electron interaction lifts the degeneracy of the single-electron energy spectrum and makes this spectrum continuous. It turns out[32,33] that the in-plane magneto-transport can still be described, even without finding elementary excitations in the correlated many-electron system, in terms of an easily calculated force on an electron from other electrons. The resulting non-monotonic magnetoconductivity obtained from this many-electron theory for not too strong $B_\perp$ is in complete qualitative and quantitative agreement with the detailed experimental data obtained by Lea, et al.[26,27] (see Fig. 6). Even the data for very strong magnetic fields have been recently understood as due to many-electron effects.[34]

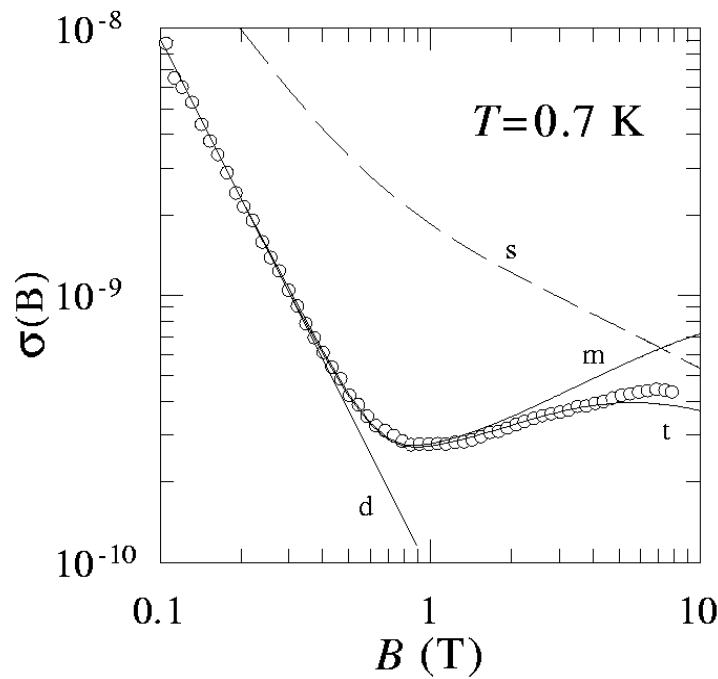

Fig.6. The magnetoconductivity $\sigma(B_\perp)$ of a correlated 2D electron system on helium for the density $n=0.55\times10^8$ cm$^{-2}$ (from Ref. 27). Circles: experimental data. Theory lines show: $m$, the many-electron conductivity in the range of classically strong and quantizing (but not too strong) fields $B_\perp$, with no adjustable parameters; $d$, the simple Drude asymptotics of the conductivity for classically strong fields (which in this case is due purely to many-electron effects); $s$, single-electron theory which ignores the electron-electron interaction, and $t$, the theory which takes into account that the electron-ripplon coupling becomes strong for strong magnetic fields.

On the whole, electrons on helium are an extremely rich and clean system, which displays various many-body effects. Many of these effects have been explored theoretically and experimentally, and are by now fully understood. In particular, we fully understand how electrons are coupled to the outside world, and we know that this coupling is very weak for low temperatures.

## 3. Making the quantum computer

### 3.1. Qubits, single-qubit gates, and decoherence effects

The above discussion suggests[19] that we can use the lowest two hydrogenic levels of an individual electron as a convenient qubit with easily manipulated level spacing and with a state which can be selectively changed, as we now discuss, by the application of resonant microwave field.

To make controllable qubits we can pattern the electrode beneath the helium surface, as shown in Figure 7, with features spaced by a distance $d$. Each micro-electrode (micro-dot) can be made to trap one electron. A trapped electron finds itself in an external potential whose detailed shape depends, of course, on the precise geometry and on the voltages $\mathbf{V}_n$ on the electrodes. The potentials $\mathbf{V}_n$ create both out-of-plane and in-plane electric fields on the electrons.

The out-of-plane field $\mathcal{E}_{\perp n}$ Stark-shifts the hydrogenic energy levels, changing the inter-level spacing. For the geometry in Figure 7 where the electrode size $d$ and the distance to the electrode from the electron are of the same order, the change of the field $\delta \mathcal{E}_{\perp n}$ is related to the change of the potential by $\delta \mathcal{E}_{\perp n} \sim \delta \mathbf{V}_n / d$. Correspondingly, the shift of the Bohr frequency of the $1 \rightarrow 2$ transition is $\sim e r_B \, \delta \mathbf{V}_n / d \, \hbar$. For $d \approx 0.5$ μm this gives $\approx 1$ GHz per 1mV of the potential increment.

A single-qubit gate is made by tuning, with micro-dot potential, the Bohr frequency to the frequency of the externally applied microwave radiation. It is well known that when a resonant field (RF) is applied to a two level system, its wave function begins to change at a rate set by the Rabi frequency which for our single qubit is quite accurately given by

$$\Omega = |\, e \, E_{\mathrm{RF}} \, \langle 2 \,|\, z \,|\, 1 \rangle \,| / \hbar \qquad (4)$$

where $E_{\mathrm{RF}}$ is the field amplitude. If the RF field is a pulse of duration $T_{\mathrm{RF}}$ and if $\Omega T_{\mathrm{RF}} = \pi$, then the system, having started in the ground state will be brought into the upper state. For $\Omega T_{\mathrm{RF}} = \pi/2$, the resulting state is an equally weighted superposition of lower and upper states. For fields of an amplitude $E_{\mathrm{RF}} \cong 1$ V/cm$^{-1}$ the Rabi frequency $\Omega \cong e E_{RF} r_B / \hbar$ is about $10^9 s^{-1}$. Thus even if we ignore the strong (see below) effect of electron confinement on the lifetime of the excited state due to ripplon-induced inter-level transitions $T_1$ (2), we could have $\Omega T_1 \geq 10^4$.

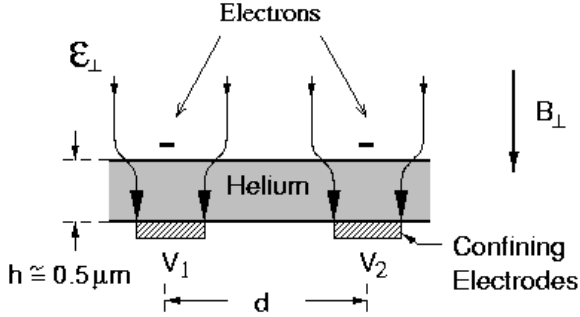

Fig. 7. Electrons on helium above a patterned substrate. The rough dimensions ($h \cong d$), the shape of the field lines, and the gate voltages are included. The average distances of the electron from the helium surface in the hydrogenic states |1> and |2> are ≈11 nm and ≈ 45 nm, respectively, for small pressing field $\mathcal{E}_{\perp}$.

Both the rate $1/T_1$ and the decoherence rate $1/T_2$, which are already small for free electrons, can be further significantly reduced by the effect of electron confinement. The in-plane confining potential arises because of the spatial pattern of the underlying electrode. If the sizes of the micro dots in Figure 7 are of the same order as the inter-dot distance $d$ and the distance to the electron layer $h$, the confinement gives rise to a set of in-plane quantum energy levels with a spacing $\hbar \Omega_{\parallel} \sim \hbar (e^2 / m_e d^3)^{1/2}$. We note that this spacing is not only fairly sensitive to the geometry of the electrodes, but also to the difference $\delta V_{nm}$ between the potentials of different micro dots. For chosen $d$, the above estimate gives $\hbar \Omega_{\parallel} \cong 0.3$ K.

In-plane quantization would suppress one-ripplon induced decay from the excited to ground hydrogenic state, if we were dealing with a single qubit. This is true because the discrete ladder of states connected with the ground and excited hydrogenic states are not commensurate by energies comparable to $\hbar \Omega_{\parallel}$. This means that the decay requires emission of a ripplon with energy $\cong \hbar \Omega_{\parallel}$ which is much higher than the energy of a ripplon whose wave vector is roughly equal to the reciprocal lateral confinement distance. However the presence of many electrons introduces additional complications. We know from our earlier discussion that electrons on the patterned electrode must form a type of pinned Wigner solid. The in-plane electronic excitation spectrum of such a solid is continuous, although in contrast to the spectrum of the unpinned solid in Figure 3, vibrational modes start for $k = 0$ not from zero frequency, but from $\Omega_{\|}$. Still the continuity of the many-electron spectrum, allows (at least energetically) a decay process with emission of one ripplon and several plasmons of the pinned Wigner solid.

We can further suppress this one-ripplon process, if we apply a strong perpendicular magnetic field $B_{\perp}$, see Figure 7. The magnetic field confines the electron to a length $\ell = \left(\hbar c / e B_{\perp}\right)^{1/2}$ and opens up gaps in the in-plane excitation spectrum of the many electron system, i.e. the spectrum now consists of a discrete ladder spaced by $\hbar \omega_c$ and broadened by $\omega_{ZB} \cong \left(\dfrac{2\pi e^2}{d^3 m_e}\right) \dfrac{1}{\omega_c}$ (if all dots are occupied). For $d = .5\mu$ and $B_{\perp} = 1.5$ Tesla, we have the magnetic length $\ell = 210$Å, $\omega_c = 2$K and $\omega_{ZB} \cong .4$K. One ripplon decay involves emission of ripplons with wave vectors limited by $k_{\ell} \cong \ell^{-1}$ and correspondingly, with frequencies smaller than $\omega_{\ell} \equiv \omega (k_{\ell}) \approx 0.4 \times 10^{-2}$ K. This makes it impossible to conserve energy in a one-ripplon decay, unless it is accompanied by emission of many phonons of the Wigner solid. Respectively, the lifetime $T_1$ becomes extremely long.

The dephasing rate of a confined electron is determined by a kind of quasi-elastic scattering of thermally excited ripplons, which is similar to the single-electron dephasing rate in quantizing magnetic fields discussed above.[28] For a confined electron the corresponding rate can be roughly estimated as

$$\frac{1}{T_2} \cong R \left(\frac{\delta_T}{r_B}\right)^4 \left(\frac{r_B}{\ell}\right)^8 \frac{R^3}{\hbar^4 \omega_{ZB}^2 \, \omega_{\ell}} \quad . \tag{5}$$

For $T = 10^{-2}$ K this gives an extremely long relaxation time $T_2 > 10^{-4}$ sec, with a ratio of the working frequency to the relaxation rate $\Omega \, T_2 > 10^5$ (a complete theory will be presented elsewhere).

To finish the discussion of the single qubit operation we note that, when a resonant microwave pulse is applied to a single confined electron to change its hydrogenic state, the presence of ripplons influences the electron resonant response in a non-trivial way. The shape of the absorption spectrum as a function of microwave frequency will have a sharp no-ripplon peak with width determined by relaxation processes, a one ripplon side band on the low-frequency side (anti-Stocks), and a broad plasmon-ripplon side band on the high-frequency side. The relative intensity of the side bands compared to the zero-ripplon peak is[19] $G \cong C_G \dfrac{R^2}{\omega_\ell^2} \dfrac{\delta_T^2 \, r_B^2}{\ell^4}$, with $C_G \sim 10^{-2}$. For T = $10^{-2}$ K, $G < .01$, and the intensity is almost all in the zero ripplon line (the parameter $G$ characterizes the strength of the electron-ripplon coupling). We note that the side bands in the absorption spectrum do not give rise to real decoherence effects. Their effects will be discussed elsewhere.

By applying different bias voltages $\mathbf{V}_n$ to individual qubits, their transition frequencies can all be independently tuned away from each other. To change the state of a targeted qubit, we apply a pulse of microwave radiation, and by varying the voltage on this qubit (on the time scale of $10^{-9}$s) we could sweep its transition frequency through the microwave frequency. With appropriately chosen sweeping rate and pulse duration, we can achieve any superposition of the ground and excited states of that qubit.

### 3.2. Two-qubit gates and the electron-electron interaction

A significant progress in conceptual mathematical discussion of quantum computing was achieved when it was shown that a "universal" quantum computer can be made of a string of qubits controlled, in addition to one-qubit gates, by two-qubit gates able to perform certain types of unitary operations.[35] A combination of these gates allows any unitary transformation to be performed on an arbitrarily large number $N$ of interacting qubits, i.e. in a $2^N$–dimensional Hilbert space of the corresponding quantum system. This means that the computer scientists encourage us to think about pairing qubits in turn, in isolation from the rest of the system, and manipulating them in a few fundamental ways. However, in real physical systems interactions are not easily turned on and off. Therefore we would argue that the manipulation of interactions in the whole interacting many qubit system, and the final measurement or collapse of the wave function must be thought of as part of the ultimate computation process. In fact, the possibility to achieve "computational universality" with two or more inputs has been discussed in literature[36].

In our computer we may pattern the initial electrodes in some arbitrary fashion, i.e. spacing, geometry, which might mean relatively isolated lines of qubits or perhaps 2D arrays, etc. We may place our electrons on those qubits, and probably arrange for some to be empty while others were full in some assigned fashion. These qubits then, having been slowly cooled, would all be in their ground state. By applying an appropriate pressing field $\mathcal{E}_\perp$ we can arrange for the level spacing to be such that transitions to the $m$ = 3 excited state were not excited. However, the qubit-qubit interactions, by virtue of the charge on the electrons, are always there. The important ones, as regards the $z$-motion, are dipole in character, and are given by $\frac{1}{2} \sum_{n \neq m} V_c \left( z_n, z_m \right)$ with

$$V_c \left( z_1, z_2 \right) \approx e^2 z_1 z_2 / d_{12}^3, \tag{6}$$

where $d_{12}$ is the distance between the electrons 1,2. We note that the interaction (6) cannot be turned off.

What we do next is to apply intense microwave (~100 GHz) pulses and relatively slow time dependent (nanosecond time scale) voltages $V_n$ to the individual electrodes. By varying these voltages and, respectively, the splitting between ground and excited states of the trapped electrons, we can not only tune qubits to resonance with the field, but in effect vary the dimensionless qubit-qubit couplings, i.e. the ratio of the Coulomb coupling to the mismatch of the transition frequencies. In the end we must read out in some fashion some features of the final wave function. The whole thing, getting to the final wave function, plus the readout must be considered as an integral part of the computation.

To make things a bit more specific and to relate them more closely to the now standard language of quantum logic gates, let's consider two neighboring qubits, with the interaction (6). At $d_{12} = d = 0.5$ μm, the Coulomb interaction (6) has two important features: (i) it effects a state dependent shift of the energy of a neighbor by about $10^{-2}$K; and (ii) it allows for the coherent resonant energy transfer of excitation from one electron to another. Suppose we start with two weakly coupled non-resonant electrons (see Fig. 7 with $V_1 \neq V_2$), one in the ground and one in the excited state (or a superposition of the ground and excited states). In order to perform the so-called swap operation, we apply a triangular shaped pressing field, i.e. increase and then decrease $V_1$. For example, this may be a linear ramp up - ramp down, with a time constant of order $10^{-9}$sec. Keeping the system in resonance for a variable time allows us to go from the wave function $|\downarrow\uparrow>$ to a wave function $|\uparrow\downarrow>$ (swap gate) or more generally, to $\cos\alpha\,|\downarrow\uparrow>+i\sin\alpha\,|\uparrow\downarrow>$ with an arbitrary $\alpha$ (the states $|\downarrow>$ and $|\uparrow>$ correspond to the ground and first excited electron state, respectively). The swap frequency is $e^2\,|<1|z|2>|^2/\hbar\,d^3 \approx 3\times10^8\,\text{sec}^{-1}$. XOR gate can also be implemented in a simple way. A detailed discussion of the gate operation for electrons on helium is given in the accompanying article by Lea *et al.*[37]

If we make a reasonable assumption that, during the calculation, the electrons may only be found in the ground and excited states $|\downarrow>$ and $|\uparrow>$ or their superposition, we can describe the electron Hamiltonian in terms of the spin one half operators $s_\alpha^n$ where $\alpha$ enumerates the spin projections. From Equation (6), the Hamiltonian of qubits can be written as

$$H = \sum_n \left[\varepsilon_n(t)\,s_z^n + F(t)\,s_x^n\right] + \frac{1}{2\hbar^2}\sum_{n\neq m}\left[A_{nm}\,s_z^n\,s_z^m + B_{nm}\,s_-^n\,s_+^m\right]. \qquad (7)$$

Here, the first term describes the one-qubit part, with $\varepsilon_n$ being the frequency of the $1\rightarrow 2$ transition controlled by the micro-dot potential $\mathbf{V}_n(t)$, and $F(t)$ being the time-dependent microwave field multiplied by $e<1|z|2>/\hbar$. The second sum in Equation (7) describes the qubit interaction, with $A_{nm} = e^2\left[<1|z|1>-<2|z|2>\right]^2/d_{nm}^3$, and $B_{nm} = 2e^2\,|<1|z|2>|^2/d_{nm}^3$. The Hamiltonian describes how our AQC will be working.

### 3.3. Other sources of noise

The interaction between electrons Eq. (6) will of course have small correction terms, which will depend on the in-plane displacements $\delta\mathbf{r}$ of the electrons from their equilibrium position, i.e. $\delta V\left(\mathbf{r}_1,\mathbf{r}_2,z_1,z_2\right) \sim V_c\left(z_1,z_2\right)\left(\delta\mathbf{r}_1-\delta\mathbf{r}_2\right)^2/d_{12}^2$. Such terms couple the hydrogenic states to the in plane degrees of freedom of the two carriers. These vibrational degrees of freedom have an energy scale characterized by $\omega_{ZB}$ . They are easily suppressed at the $10^{-5}$ level because $(\delta r_1 - \delta r_2)^2 / d^2 < 3 \times 10^{-3}$.

Other types of very weak dissipative processes can be present, but are not serious. More specifically; 1) spontaneous radiative emission is negligible, 2) non-radiative decay due to pair creation in the electrodes would be significant but can be completely suppressed by using superconducting electrodes with a gap greater then the transition frequency, and 3) voltage noise. However, assuming that the potential is applied using a transmission line with a resistance of 50 Ohms, having a noise temperature of 10K, then the capacitative plates (our electrodes) will be subject to a noise voltage of roughly $10^{-10}$ V/$\sqrt{\text{Hz}}$ . For the geometry we discuss, the corresponding fluctuations of the frequency of the $1 \rightarrow 2$ transitions are $10^2 \sqrt{\text{Hz}}$ . This gives the decoherence time $10^{-4}$s, which is acceptable.

### 3.4. Read-out

In order to read out the wave function at some time, $t_f$ , when the computation is completed, we apply a reverse field $\mathcal{E}_+$ to the capacitor which would, if it were large enough, remove all the electrons from the helium surface. However, if the reverse field at the electron layer is not too strong, the electrons see a potential that has a maximum similar to that shown in Fig. 8.[38] For this potential the electron tunneling rate depends exponentially on the barrier height $V_{Bm}$. The electrons will clearly leave the surface in some time, $t_m$ which is a strong function of the state $m$. It is possible to set the field $\mathcal{E}_+$, so that if one waits for some reasonable time, say one microsecond, only the excited electrons will come off. They will arrive at the plate (anode) with some modest kinetic energy, i.e. tens of electron volts. If the anode is part of a low temperature channel plate arrangement with a spatial resolution of 1μm or less, those electrons may be imaged and a "picture" taken of the wave function.

Of course each time we do this operation we get a different result. The probability distribution gives us the "answer" to the problem we were solving. In this case it is clear that the results we get will very much depend on the exact time dependence of the reversing potential as well as the ultimate size of $\mathcal{E}_+$. As the potential is reversed, the wave function will continue to evolve in time. This is true even though we are no longer applying RF pulses or sweeping the gate potential, since we are generally not in an eigenstate of the system. Then tunneling is on all the time. If the reversal time is short compared to the tunneling time, then the density matrix of the electrons localized on helium surface $\rho(t)$ is described by the equation

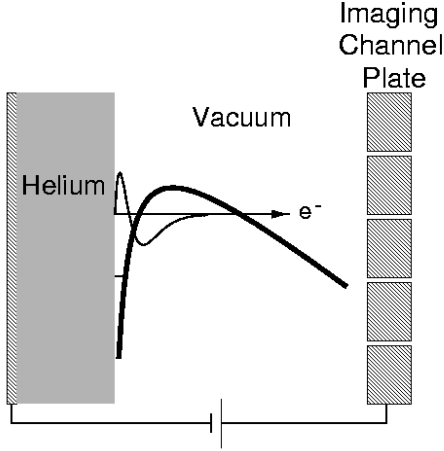

Fig. 8. The read out configuration. The potential leading to tunneling in the presence of a reverse field is shown along with a schematic of the wave function of the excited state and the position sensitive channel plate.

$$\frac{d\rho}{dt} = i\hbar^{-1}[\rho, H(t)] - \Theta(t - t_f)(2\hbar^2 t_\uparrow)^{-1} \sum_n [s_+^n s_-^n \rho + \rho s_+^n s_-^n], \qquad (8)$$

where $H(t)$ is the Hamiltonian of interacting qubits (7), $\Theta(t)$ is the step function, and $t_\uparrow$ is the tunneling time from the excited state [note that, because of the tunneling, the number of localized electrons is not conserved]. From Eq. (8) one can find the probabilities for the system to occupy a given state, for example one of the $2^N$ non-interacting basis states, and relate the result of the calculation to the site distribution of the electrons that will have tunneled from the system.

Another possibility is to measure directly the change of voltage on the patterned electrodes, which results from interstate electron transitions and is about $er_B/d^2 <\sim 0.1$ mV. This is equivalent to attaching single electron transistors (SET) to the control electrodes. This possibility will be discussed by M. Lea *et al.* [37] One of its advantageous features is that there is no need in re-charging the system after each measurement.

## Conclusions

Besides doing an actual computation, there are many exciting physically interesting problems, which could be studied using our analog system. For example, making the energy levels randomly disordered in a controlled way, and exciting only one electron, we could study the single-particle localization and investigate such features of quantum chaos as level statistics and the structure of the wave function. Putting in more energy, for example two excitations, we could study the localization and chaos in a system of two interacting excitations which would for $N$ qubits involve, if we did it numerically, the diagonalization of a $[N(N-1)] \times [N(N-1)]$ matrix. Putting the system into some highly excited state we might try to map its motion in Hilbert space, decide how often it returns to its original configuration, how the distribution over energies evolves in time, and therefore solve the quantum Fermi-Pasta-Ulam problem. We could go to the dilute limit, i.e. have non-interacting qubits and investigate the problem of driving a single, in this case 1D hydrogen atom into very highly excited states whose structure could be controlled with voltage bias. This is the problem of quantum chaos in Rydberg states of a new type.

The system of two-dimensional electrons on helium is unique in the context of large AQC systems. It is scalable, easily manipulated, and has perfectly acceptable decoherence times. It has been carefully investigated theoretically and experimentally, and there seems to be no existing technological barriers present for making an AQC using it.

The authors would like to thank Arnold Dahm, John Goodkind, and Michael Lea for many interesting conversations and correspondence regarding the experimental feasibility of these suggestions and for agreeing to actually launch experiments on such systems. We would also like to acknowledge many discussions with Mike Andrews and Allen Mills about quantum computing.

MID acknowledges financial support from the Center for Fundamental Materials Research at MSU.